\documentclass[aip,jap,amsmath,amssymb,reprint,superscriptaddress]{revtex4-1}
\usepackage{graphicx}
\usepackage{amsmath}
\usepackage{dcolumn}
\usepackage{bm}
\usepackage{bbold}
\usepackage{color}
\usepackage{setspace}

\usepackage{caption}
\usepackage{subcaption}

\begin{document}

\title{Anomalously Strong Nonlinearity of Unswept Quartz Acoustic Cavities at Liquid Helium Temperatures}

\author{Maxim Goryachev}

\author{Warrick G. Farr}

\author{Eugene N. Ivanov}

\author{Michael E. Tobar}
\email{michael.tobar@uwa.edu.au, Tel.: +61-8-6488-3915, Fax: +61-8-6488-1235}
\affiliation{ARC Centre of Excellence for Engineered Quantum Systems, University of Western Australia, 35 Stirling Highway, Crawley WA 6009, Australia}

\date{\today}


\begin{abstract}
We demonstrate a variety of nonlinear phenomena at extremely low powers in cryogenic acoustic cavities fabricated from quartz material, which have not undergone any electrodiffusion processes. Nonlinear phenomena observed include lineshape discontinuities, power response discontinuities, quadrature oscillations and self-induced transparency. These phenomena are attributed to nonlinear dissipation through a large number of randomly distributed heavy trapped ions, which would normally be removed by electrodiffusion. A simple mean-field model predicts most of the observed phenomena. In contrast to Duffing-like systems, this system shows an unusual  mechanism of nonlinearity, which is not related to crystal anharmonisity. 
\end{abstract}

\maketitle

\section{Introduction}

Quartz Bulk Acoustic Wave (BAW) cavities have recently been shown to have unprecedented quality factors both at liquid helium and milli-Kelvin temperatures\cite{Goryachev1,quartzPRL}. 
It has been demonstrated that highly efficient crystal clamping techniques enable such resonators to operate in a regime in which their quality factor is primarily limited by the material acoustic losses\cite{1537081}. In particular,
 phonon-phonon interaction losses described by the Landau-Rumer theory\cite{landaurumer1}. In this regime, the acoustic phonons of the driven modes are scattered by thermal phonons of the material modes due to the crystal anharmonicity. Despite the fact that all Taylor series terms of the Hamiltonian representing the lattice anharmonicity could potentially take part in the loss mechanisms, only the linear losses resulting from the lowest (three-phonon mixing) terms\cite{Maris1971} have been detected in extremely high-$Q$ BAW cavities. So, although the crystal anharmonicty is involved in the loss mechanism, the mode behaviour remains linear until the point where self interaction between acoustic phonons becomes pronounced. It is this mixing between acoustic phonons that leads to the well known nonlinear effects in BAW devices, which are well-described by a Duffing oscillator model both at room and cryogenic temperatures\cite{nosek, SunFr}. 
 
Another way acoustic cavities can demonstrate anomalous nonlinear behaviour is due to a higher degree of disorder resulting from various sorts of defects. Provided that the dissipation due to the clamping mechanisms are minimised, disordered inclusions can lead to significant macroscopic effects. This work demonstrates a variety of such phenomena. We relate this behaviour to heavy randomly-distributed ions trapped in potential wells near substitutional defects. These trapped ions couple to acoustic waves in the crystal by causing nonlinear dissipation as the mechanical vibration of the crystal lattice cells causes motion of these ions in their  anharmonic potential wells. We report results of a unique low loss acoustic system coupled to a large number of nonlinear dissipation elements. All other systems reported have either had too high clamping losses, which have degraded the acoustic Q, or too low concentration of impurities to enter this regime. 

\section{Measurement Technique}

Ultra-high quality acoustic cavities were typically characterised by the impedance analyser method~\cite{Galliou:2008ve, galliou:091911, Goryachev1}. This method has a significant limitation due to the limited power range of the measurement apparatus. This limitation can be a serious problem for highly nonlinear low loss systems. Thus, in this work, we utilised an RF signal splitter in order to characterise the cryogenic BAW cavity at  extremely low powers.


The measurement setup is shown in Fig.~{1}. The network analyser was set to transmission measurements. The probe signal was attenuated and injected into one of the arms of the splitter. Part of the signal was then reflected from the Device Under Test (DUT) and the output signal was taken from the other arm of the splitter, amplified and measured by the network analyser. Thus, the setup indirectly measured the reflection coefficient of the DUT at low incident powers.

The drawback of the approach was the inability to reliably calibrate the system. Thus, the measured results were only specific to the loaded device. However, the subject of this study was not precise determination of absolute characteristics, but rather a study of the sensitivity to the excitation power. For this reason, all the response results are presented in relative arbitrary units. Moreover, due to the low coupling of the studied modes, their loaded quality factor is virtually the same as the unloaded values.

\section{Unswept Quartz Acoustic Cavity}

The cavity under investigation was an SC-cut electrode-deposited bulk acoustic wave resonator made from unswept low etch channel density synthetic quartz, which was cooled to liquid helium temperatures. The device was designed to work on the shear thickness mode.  

This resonator exhibits vibrations of several of the lowest order overtones (OT) of the thickness shear-polarized modes.  In particular, at liquid helium temperatures, it is possible to excite the fundamental ($\sim20$~MHz), third ($\sim60$~MHz) and fifth ($\sim100$~MHz) OTs. The OT number denotes the number of half-waves along the plate thickness and several modes with different in-plane wave numbers are usually observed. The quality factors of such spurious modes are higher than that of the main resonance due to significantly lower coupling to the measurement (electrical) environment.  The lower coupling leads to higher loaded quality factors and for this reason, in the following study only modes with high in-plain wave-numbers were considered.  

The nonlinear effects observed in this paper are attributed to the microscopic defects of the crystalline structure that are more dense in the cavity studied in this work than in all previously characterised devices. The main difference between the acoustic cavities studied in this work and others studied earlier\cite{Galliou:2008ve, galliou:091911, Goryachev1} is the material treatment. In particular, the synthetic quartz used for manufacturing cavities in this work did not undergo the electrodiffusion (sweeping) procedure\cite{Halliburton,Martin1988}. The procedure entails clamping synthetically grown quartz bars between conducting electrodes (with the electric field density of the order of $2-3\times10^3$ V/cm) under vacuum at high temperatures ($\sim500^\circ$C) for about 10 days, and targets two types of defects (with the first being the most pronounced): 
\begin{itemize}
\item Alkali impurities (such as $Li^+$ and $Na^+$) trapped adjacent to substitutional $Al^{3+}$ ions, other point defects are substitutional $Ge$ and both silicon and oxygen vacancies\cite{weil,Martin}.
\item Extended dislocational networks with precipitated impurities that form an etch channel.  
\end{itemize}
Under the electrodiffusion process alkali ions are replaced either by hydrogen atom nuclei (protons) released by growth defects giving $Al-OH^-$, or by electronic holes coming from the anode forming $Al-h^+$ paramagnetic defects\cite{weil}. It has been demonstrated by many authors that sweeping has a drastic effect on improvement of the quartz acoustic losses at room temperatures ( for example, see~\cite{Martin1988,Euler1982} and references therein).

It has also been reported that electrodiffusion of quartz significantly reduces the etch tunnel density\cite{vig77} \cite{Martin1983,Gualtieri}. However, as it is suggested by X-ray topography\cite{Martin1988,lipson}, the process substitutes the alkali ions in dislocation networks with protons but leaves the dislocation networks intact. Thus, the total effect is again reduced to ion substitution. Nevertheless, the material under study here is still qualified as a 'low etch channel' (tubular cavities generated along linear defects in a crystal), which was reduced using an alternative technique leaving this parameter comparable to that of the previously used quartz systems. Thus, the main difference between the unswept quartz used in this study to that of other studies is the presence of heavy potentially trapped alkali impurities such as $Na^+$, $Li^+$, $K^+$ (of up to $5$ppm) and near substitutional $Al^{3+}$ (of up to $10$ppm), as well as alkali ions in dislocation networks substituted by light holes and protons. 

\section{Measurement Results}

The modulus of the acoustic cavity reflection coefficient was measured as a function of incident power on the crystal {and the detuning frequency $\Omega$ (difference between the resonance and the pump signal frequencies)}, with results shown in Fig.~{{2}} for anharmonics of the fundamental mode, Fig.~{{3}} for anharmonics of the third overtone and Fig.~{{4}} for anharmonics of the fifth overtone. 
These modes exhibit following types of nonlinear effects; 
\begin{itemize}
\item frequency sweep discontinuity (also referred as 'jump' phenomenon), which manifested as a abrupt response moving away from the resonance with increasing incident power;
\item frequency-power dependence, seen as frequency shift with increasing power;
\item power sweep discontinuity, observed as an abrupt change of the modulus of the reflection coefficient for a given frequency with only a small increase of incident power;
\item oscillatory instability regions, which appear as areas with rapidly varying reflection due to relaxation oscillations of the response;
\item self-induced transparency, which was seen as an avoided crossing with incident power causing the doubling of the frequency response extrema at given values of incident power.
\end{itemize}
These anharmonics were chosen due to their higher sensitivity to the incident power. Other anharmonics exhibit similar behaviour but at higher power levels, or are much less sensitive. The fundamental modes for each OT exhibit lower quality factors. Quality factors of investigated anharmonics vary in the range $1-30\times10^6$.

Figures {{2}},~{{3}},~{{4}} show variety of nonlinear effects, the most noticeable of which are self-induced transparency (Fig.~{{5}}), line shape discontinuities (Fig.~{{6}}), relaxation oscillations (Fig.~{{7}}), etc. Relaxation oscillations happen at the border of two regions separated by a discontinuity. The frequency of oscillations depends on how close the pump signal is to one or another regions. 

The demonstrated results are repeatable provided that the incident power on the crystal does not exceed a certain threshold level. However, once exceeded the behaviour is again repeatable only if the crystal is cycled back to room temperature and re-cooled again. Similar behaviour has been observed for a second device of the same type.  

An important fact is that all the nonlinear effects are observed at considerably low powers. The BAW cavities studied previously~\cite{Galliou:2008ve,galliou:091911,Goryachev1} while exhibiting much higher quality factors, demonstrate nonlinear effects only at incident powers above $-35$~dBm. Moreover, all these effects are well explained by crystal anharmonicity and temperature variation and fit well into the Duffing model~\cite{SunFr}. In contrast, in this work none of the demonstrated nonlinear effects fall into this model and thus cannot be explained by the crystal anharmonicity. In addition, the fractional frequency temperature sensitivity of the devices studied in this work was of the order $10^{-8}$~K$^{-1}$, which is comparable to that of the previously studied devices~\cite{SunFr} and rules out temperature effects at such low powers.


The temperature dependance of two modes is show in Fig.~{{8}}. Both density plots show that there is a temperature with maximum mode coupling. The maximum coupling corresponds to the minimum of the reflection coefficient obtained at different temperatures. Although it is less visible for the plot (a) where the maximum is achieved at relatively low temperature $5$K, this behaviour becomes apparent for the OT at plot (b) where the maximum coupling temperature is $10.45$K.

The nonlinear effects demonstrated in this section cannot be referred to the typical nonlinearity of acoustical devices since none of the demonstrated effects can be modelled with the Duffng-type oscillator. The frequency-power response of this model is widely known with a corresponding well known frequency shift and bistability effect. The Duffing bistability occurs only to one side of the linear resonance frequency, and thus does not fit the experimental observations made here. Moreover, typical nonlinearities of cryogenic acoustic cavities due to lattice anharmonicity become visible at significantly high incident powers, typically in the range $-40..-30$~dBm, where nonlinear term of the equation of motion becomes comparable to the linear angular frequency term\cite{Goryachev1,SunFr}.  For the same reason the observed phenomena cannot be due to thermal effects in the cavity, since typical self-heating effects are observed at even higher dissipated powers. Moreover, the characterised modes demonstrate temperature sensitivity typical for this type of devices at these temperatures\cite{Galliou:2008ve, galliou:091911}. In this work, we relate the observed nonlinear phenomena to nonlinear dissipation in the cavity due to lattice impurities discussed in the previous section. This situation structurally corresponds to the van der Pol oscillator that has to be set stable in the origin. Indeed, since the nonlinearity comes into the model through the dissipation term, its impact has to be compared to the resonance bandwidth rather than to the absolute angular frequency. This means a much higher power sensitivity was observed in these measurements. This model is discussed in the following section.

\section{Simplified Mean-Field Model of Microscopic Nonlinear Losses}

The simplest phenomenological model of the acoustical vibration in a cavity with heavy localised defects is given by the one-dimensional forced wave equation: 
\begin{equation}
\label{A000DF}
u_{tt} = v^2u_{xx}+\Gamma u_{xtt}+\sum_i\delta(x-x_i)G_i(u_{x},u_{t})+F(t),
\end{equation}
where $u$ is the displacement at the point, $v$ is the speed of sound ($v = 3611$ m/s for the slow quasi-shear mode in quartz), $F(t)$ is the forcing term related to the piezoelectric properties of quartz. The second right hand side term of this equation represent linear unlocalized losses. The third term represent  losses due to interaction with local ($\delta(x-x_i)$) nonlinear dissipative systems. The $G_i$ functions introduce small perturbation on top of the linear losses.  
Such a model is used to explain similar behaviour of acoustic waves in resonant bars with artificially introduced defects~\cite{Fillinger1, Fillinger2}.

In this case we consider the simplest case when $G_i$ are regarded as perturbation on the solution (eigenmodes) of the linearized system. 
In the special case of one defect, wave equation problem (\ref{A000DF}) and corresponding boundary conditions are reduced to the second order differential equation with nonlinear dissipation:
\begin{equation}
\label{A001DF}
\ddot{u}+2\gamma(1+\alpha(u,\dot{u}))\dot{u}+\omega_0^2u=2\omega {f}
\end{equation}
where $f=A\cos\omega t$ is the forcing term, $\omega_0$ is the angular frequency of the characterised mode, $\gamma$ is the damping coefficient and $\alpha$ is a nonlinear function approaching zero for small $u$ and $\dot{u}$ that can be expanded in a Taylor series.
 Regardless of its simplicity, equation (\ref{A001DF}) can predict both most distinctive experimental observations: jump phenomenon and self-induced transparency.
Note that equation (\ref{A001DF}) is different from the van der Pol oscillator because it contains a positive linear loss term. By applying the perturbation approach with the averaging technique, the problem can be reduced to analysis of signal quadratures:
\begin{equation}
\label{A002DF}
\left. \begin{array}{ll}
\displaystyle \frac{d}{d\tau}U=-\gamma \Big[1-\sum_n\eta_n M^{2n}\Big] U+ \Omega \Big[1-\sum_n\xi_n M^{2n}\Big] V +A,\\
\displaystyle \frac{d}{d\tau}V=-\Omega\Big[1-\sum_n\xi_n M^{2n}\Big] U-\gamma  \Big[1-\sum_n\eta_n M^{2n}\Big] V ,
\end{array} \right. 
\end{equation}
where $M^2=V^2+U^2$ is vibration magnitude, $A$ is the magnitude of the incident signal (external forcing term), $\tau$ is a slow time representing time-variation of the signal quadratures, $\Omega = \frac{\omega_0^2-\omega^2}{2\omega}$ is a detuning frequency, $\xi_n$ and $\eta_n$ are functions of Taylor expansion coefficients of the function $\alpha$, $\gamma$, $\omega$ and $\Omega$. The system is analysed in a steady-state regime corresponding to frequency response measurements, when both quadratures are independent of time. This leads to a polynomial equation: 
\begin{equation}
\label{A003DF}
\left. \begin{array}{ll}
\displaystyle A^2= \Big[\gamma^2 \big(1-\sum_n\eta_n M^{2n}\big)^2+\Omega^2\big(1-\sum_n\xi_n M^{2n}\big)\Big] M^2,
\end{array} \right. 
\end{equation}
whose solutions give a magnitude response of the system   as a function of detuning frequency $\Omega$ and incident power $A^2$ (equivalent of Fig.~{{2}},{{3}},{{4}}).

The difference between the system reduced to (\ref{A003DF}) and the widely studied Duffing oscillator, is that the latter does not have magnitude dependent  $\gamma^2$ coefficient. The direct consequence of that is that equation possesses frequency independent nonlinear coefficients. Moreover if the $\Omega^2$ nonlinear terms can be ignored, the solutions should be symmetric around $\Omega=0$. Thus, if the system (\ref{A003DF}) has three solutions they appear as frequency sweep discontinuities ('jumps', see Fig.~{{6}}) moving away from the linear resonance with increasing power. In contrast, for the Duffing oscillator, discontinuities exists only on one side of the $\Omega=0$ line and detune away from it for increasing $A^2$, whereas our system demonstrates symmetric discontinuities detuning away from the $\Omega=0$ line on both sides for increasing $A^2$. The clear picture of this is given in Fig.~{{3}}, (d) where there are three symmetric discontinuities ('jumps'). This effects was also observed in the other anharmonics but is not shown here. 

The magnitude instabilities (Fig.~{{7}}) exist for the regions with free real solution of equation (\ref{A003DF}) where system (\ref{A002DF}) becomes unstable. This leads to van der Pol-like relaxation oscillations. 

To confirm the fact that the nonlinear loss mechanism predicts experimentally observed self-induced transparency, we solve equation (\ref{A003DF}) for the case of $\xi_n=0$ and to second order in $\eta$ by the perturbation approach. The solution of the linearized system (or zeroth-order approximation) is $M^2_{(0)}=A^2/(\gamma^2+\Omega^2)$, and can be used to calculate the first order approximation of the inverse of the absolute value of the transmission coefficient, given by:
\begin{equation}
\label{A004DF}
\left. \begin{array}{ll}
\displaystyle \frac{A^2}{M^2}=\Omega^2+\gamma^2\Big[1-\eta_1M^2_{(0)}-\eta_2M^4_{(0)}\Big]^2.
\end{array} \right. 
\end{equation}
If the frequency magnitude response of the system at certain power $A^2$ demonstrates self-induced transparency, then the magnitude of inverse of the transmission coefficient (\ref{A004DF}) will have two maxima. This could be demonstrated by the first derivative of the expression with respect to $\Omega$ having three zeros. Apart from a trivial zero $\Omega=0$, other two zeros are solutions of the following polynomial equation:
\begin{equation}
\label{A005DF}
\left. \begin{array}{ll}
\displaystyle X^5  + 2 A^2  \gamma^2 \eta_1 X^3 + 2 A^4  \gamma^2 (2\eta_2 - \eta_1^2) X^2 - \\
\displaystyle -6 A^6  \gamma^2 \eta_2 \eta_1 X  - 4 A^8 \gamma^2 \eta_2^2=0,
\end{array} \right. 
\end{equation}
where $X=\Omega^2+\gamma^2$ and thus only $X>\gamma^2$ solutions make physical sense. Using Descartes' rule of signs, it is possible to show that for certain signs of $\eta_1$, $\eta_2$ coefficients and their relations, equation (\ref{A005DF}) has solutions satisfying $X>\gamma^2$ requirement. 
It can be also shown that the corresponding equation with $\eta_2=0$ does not give physically grounded solutions. 
 This means that for $\eta_1\neq0$ and $\eta_2\neq0$, the frequency response of the system exhibits several maxima, known as self-induced transparency. 
 
\section{Conclusion}

Mechanical resonators play a very important role as instruments in the investigation of a wide variety of fundamental physics problems including quantum measurement and computation\cite{Braginsky1992}, spin detection, ultra-sensitive force detection and more. Thus, the demonstrated phenomena have potential applications in a variety of areas of physics. On one hand, the observed nonlinear effects are unwanted phenomena whose manifestation should be avoided. Thus, the study demonstrates the importance of crystal treatment for manufacturing of ultra-low loss acoustic cavities\cite{ScRep}. In particular, it demonstrates that impurities can significantly limit the cavity performance even at relatively modest level of quality factor. We have demonstrated effects of strong impurity-phonon coupling never observed before, which inevitably leads to a better understanding of impurities in crystal in solids\cite{Luthi} and their interaction with acoustical waves\cite{Lefevre}. 

On the over hand, low-driving power nonlinear systems are of the special interest in many applications of the modern physics due to their rarity, with only one another example being a Josephson junction. Such nonlinearities can be exploited to build cryogenic parametric amplifiers and other electronic systems etc\cite{supel}. In addition to that, low-loss mechanical systems represent special interest for engineered quantum applications\cite{,OConnell:2010fk}. For example, in the system  understudy is nearly in the ground state at the temperatures of $20$mK for the 5th OT, which is accessible with conventional dilution refrigerators. Thus, the strong internal mode nonlinearity may serve as a tool for exploration of nonlinear phenomena in quantum physics. In addition, the demonstrated interaction of phonons represent an emerging area of phonon cavity elastodynamics\cite{Ruskov, Mahboob}.
 
 
\begin{figure}[t!]
\centering
\includegraphics[width=3.25in]{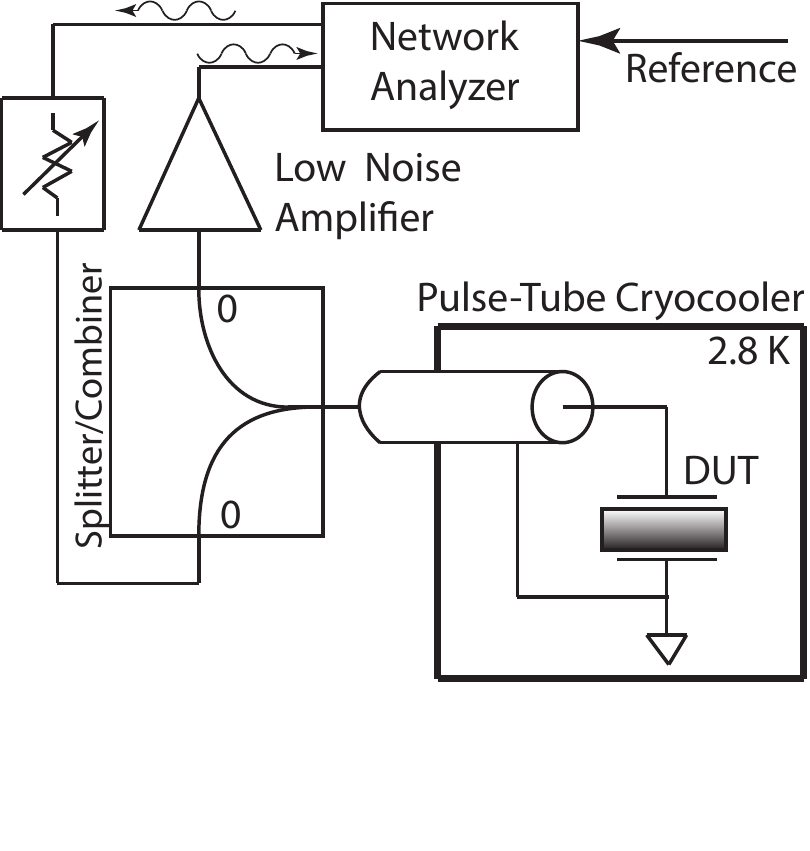}
 \caption{The measurement setup, show that the device under test (DUT) was characterised using indirect reflection measurements. The probe signal was first attenuated before the actual injection into the DUT, and then amplified on the way back. }\label{measurement}
\end{figure}

\begin{figure*}[t!]
        \centering
        \begin{subfigure}[b]{0.5\textwidth}
                \centering
                \includegraphics[width=\textwidth]{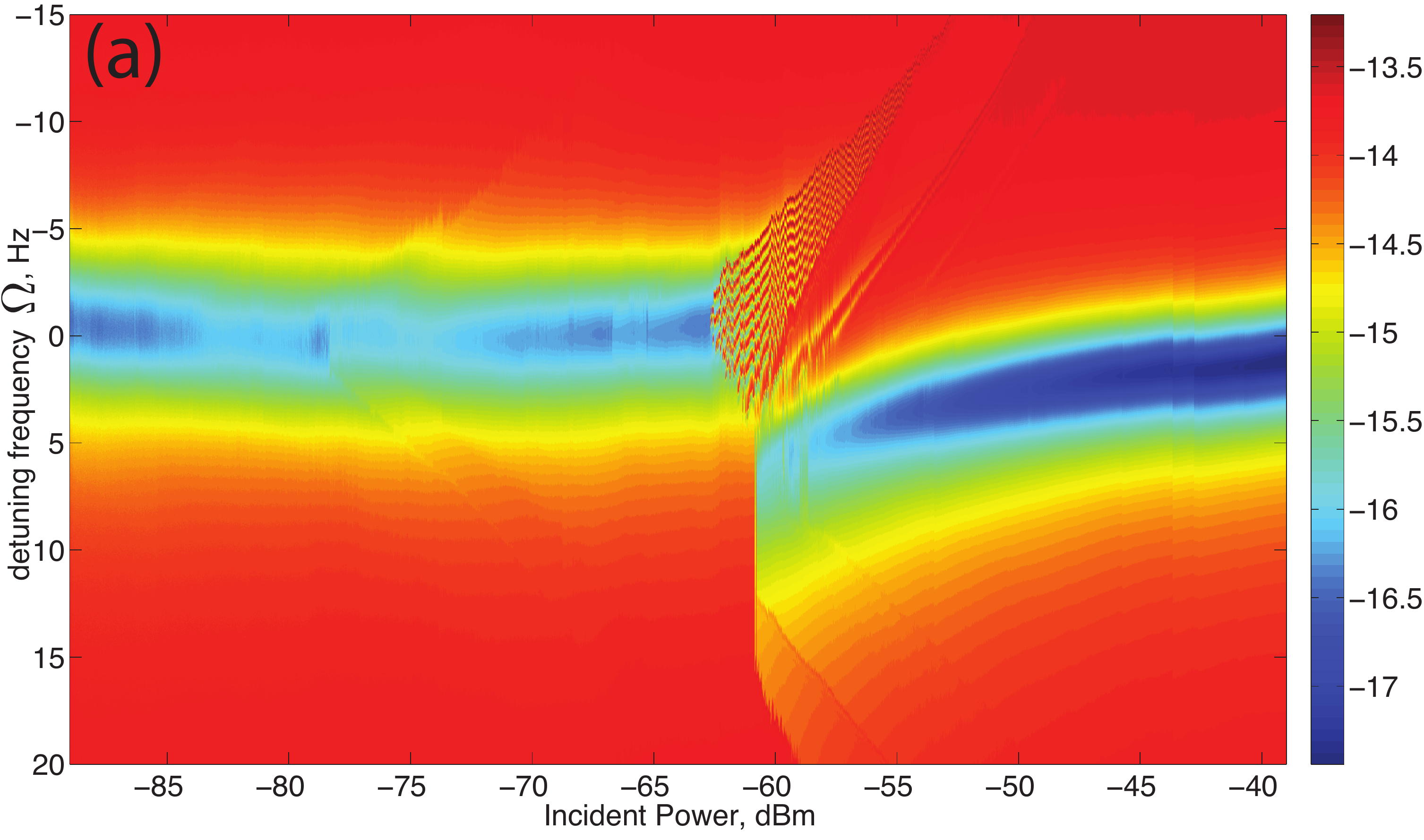}
                \label{magn1A}
        \end{subfigure}%
        ~ 
        \begin{subfigure}[b]{0.5\textwidth}
                \centering
                \includegraphics[width=\textwidth]{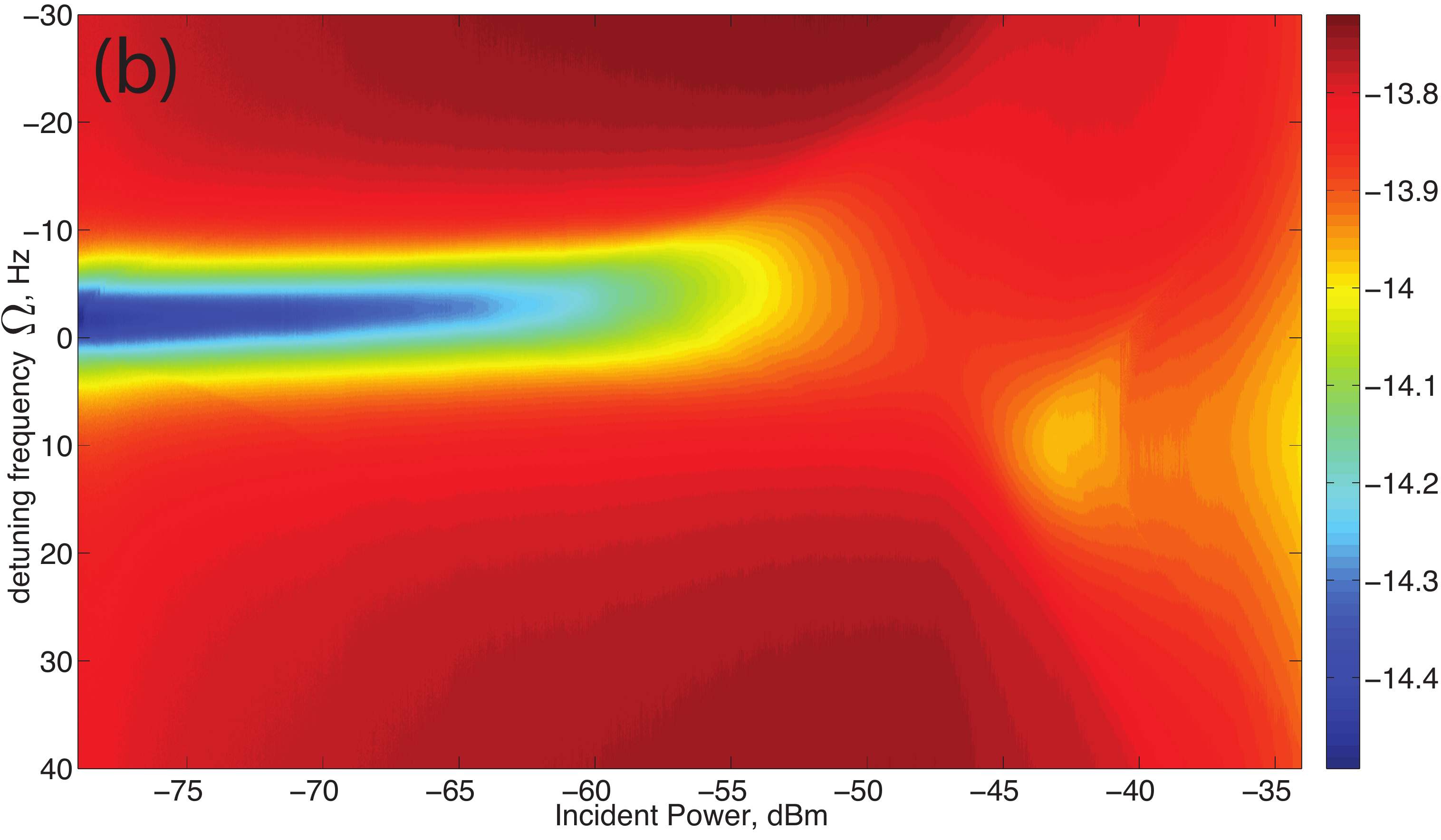}
                \label{magn1B}
        \end{subfigure}
        \caption{Power Sensitivity of some anharmonics of the fundamental mode (a: $20.04608$~MHz, b: $19.98659$~MHz) in terms of magnitude of the reflection coefficient. Here (a) exhibits multiple 'jump' and power sweep discontinuities frequency-power dependence and an oscillatory instability region, while (b) demonstrates self-induced transparency followed by radially growing jumps starting at about $-42$dBm.}\label{mags}
\end{figure*}

\begin{figure*}[t!]
        \centering
        \begin{subfigure}[b]{0.5\textwidth}
                \centering
                \includegraphics[width=\textwidth]{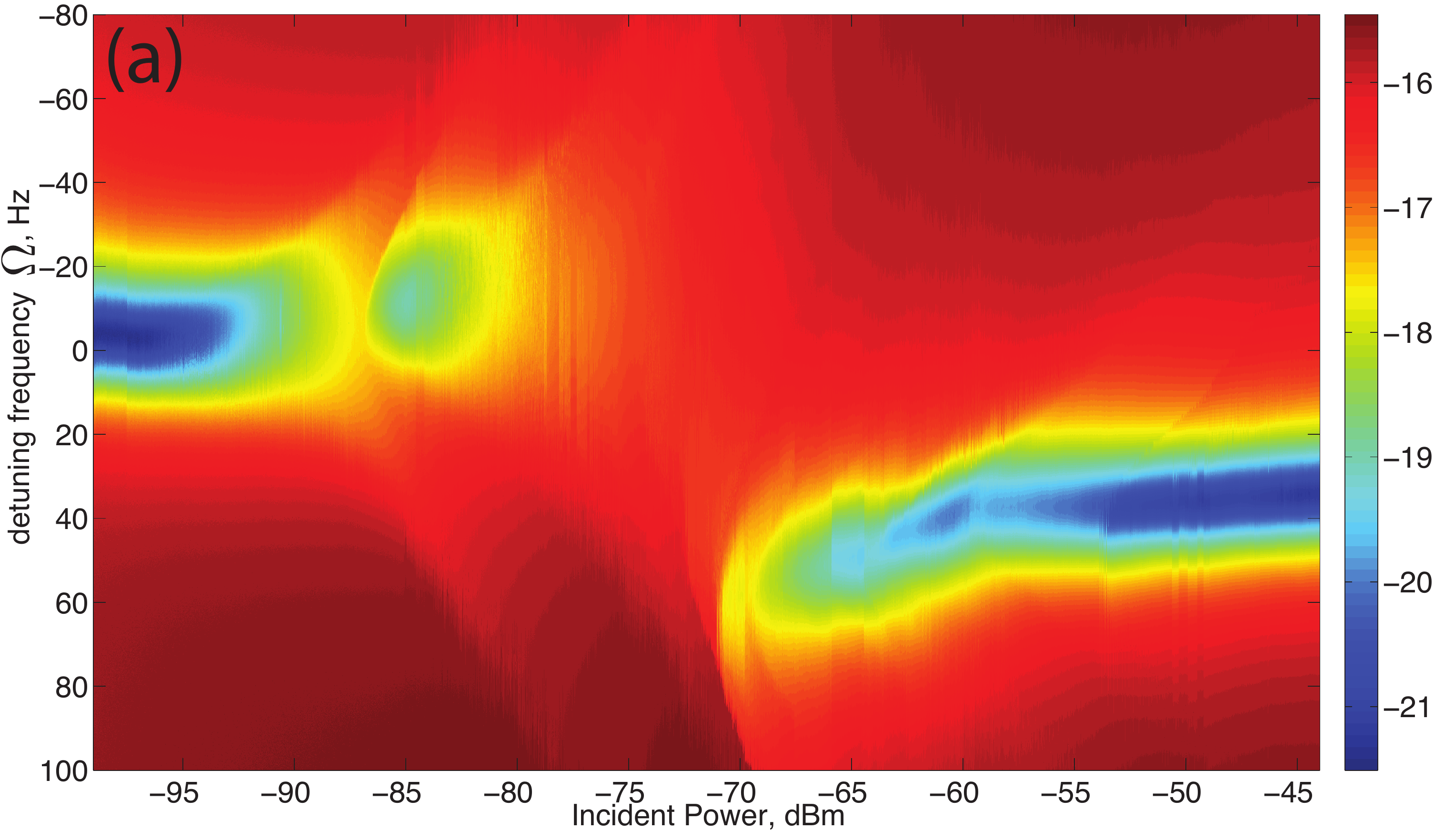}
                \label{magA13}
        \end{subfigure}%
        ~ 
        \begin{subfigure}[b]{0.5\textwidth}
                \centering
                \includegraphics[width=\textwidth]{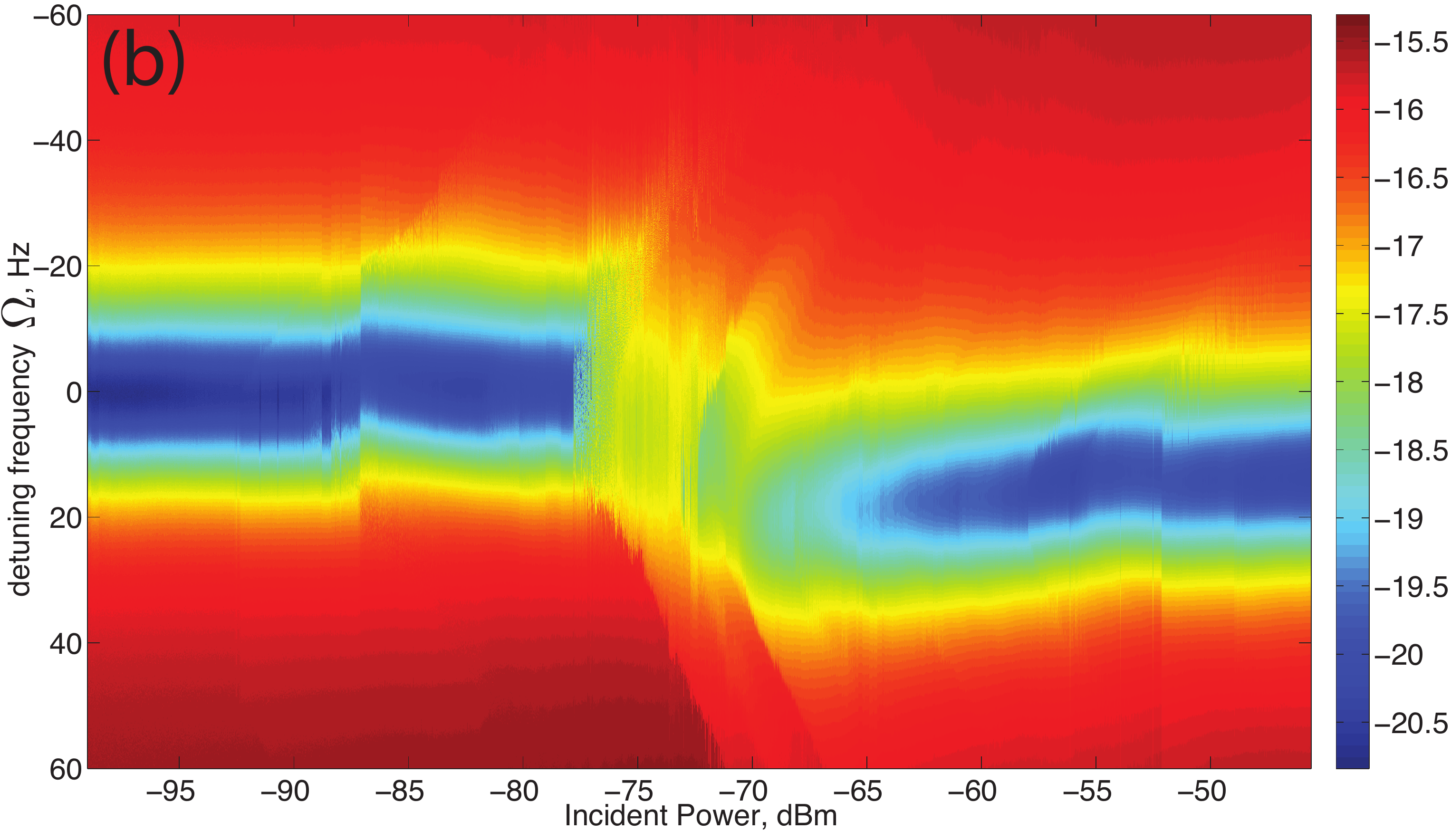}
                \label{mag31B}
        \end{subfigure}        
           
           \centering
       \begin{subfigure}[b]{0.5\textwidth}
                \centering
                \includegraphics[width=\textwidth]{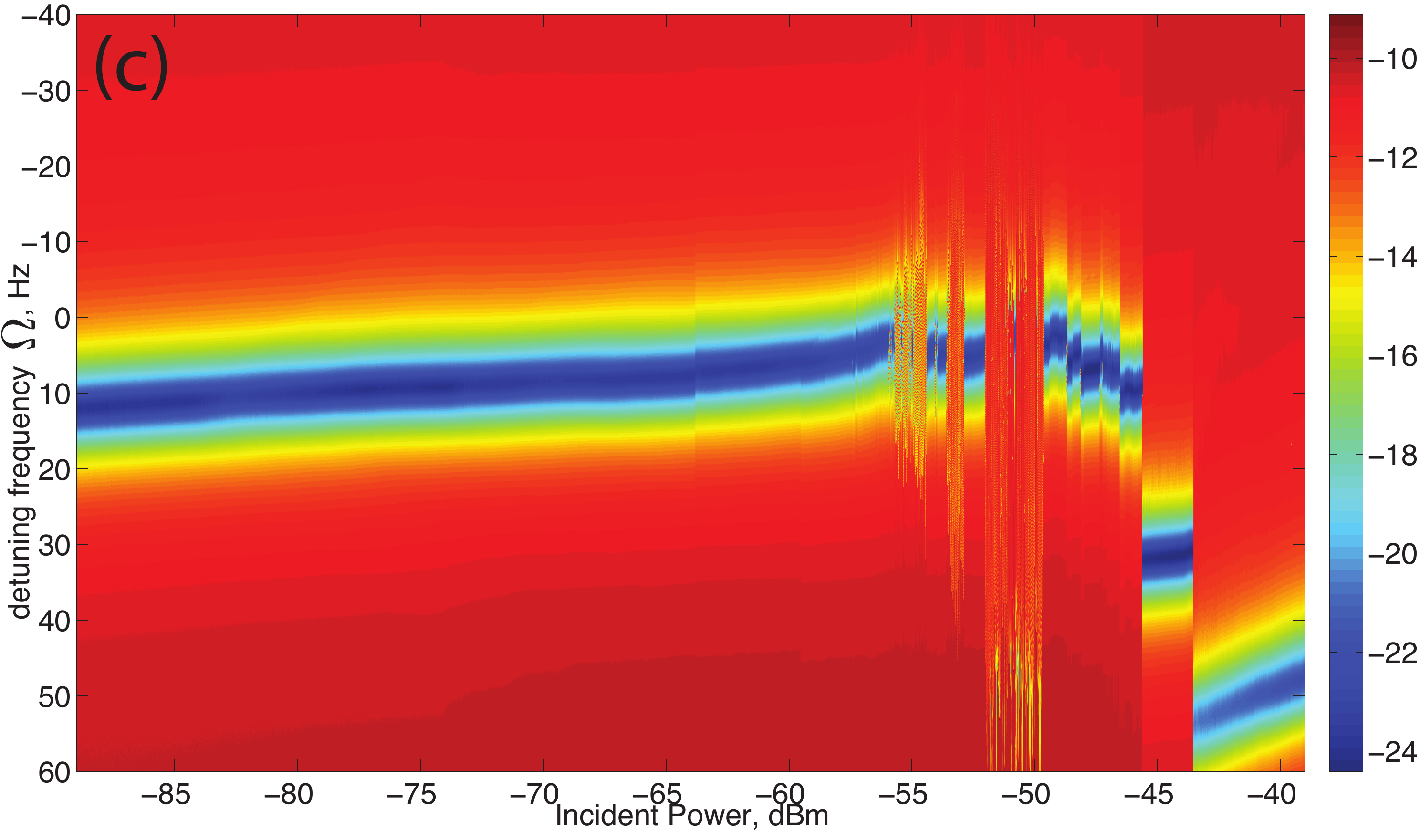}
                \label{magA3}
        \end{subfigure}%
        ~ 
        \begin{subfigure}[b]{0.5\textwidth}
                \centering
                \includegraphics[width=\textwidth]{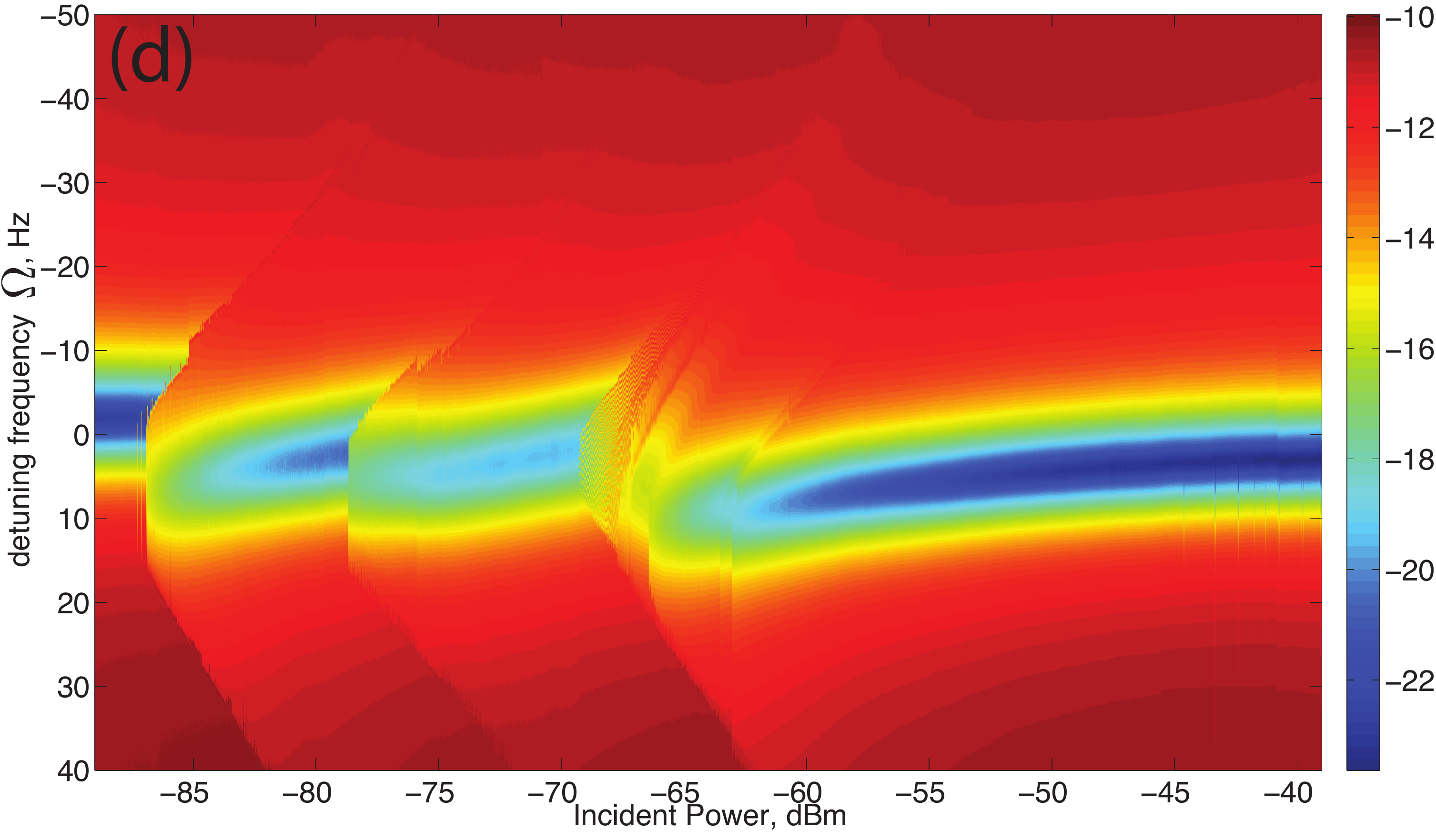}
                \label{mag3B}
        \end{subfigure}        
           \caption{Power Sensitivity of some anharmonics of the 3rd OT (a: $59.96344$, b: $59.96925$, c: $59.92972$, d: $59.94871$) in terms of magnitude of the reflection coefficient. Here (a) and (b) have self-induced transparency accompanied by frequency sweep discontinuities ('jumps'), while (c) is subject two power sweep discontinuity and relaxation oscillations and (d) shows multiple radially moving away discontinuities decorated by oscillatory instability of the high power one.}\label{mags2}
\end{figure*}

 \begin{figure*}[t!]         
           \centering
        \begin{subfigure}[b]{0.5\textwidth}
                \centering
                \includegraphics[width=\textwidth]{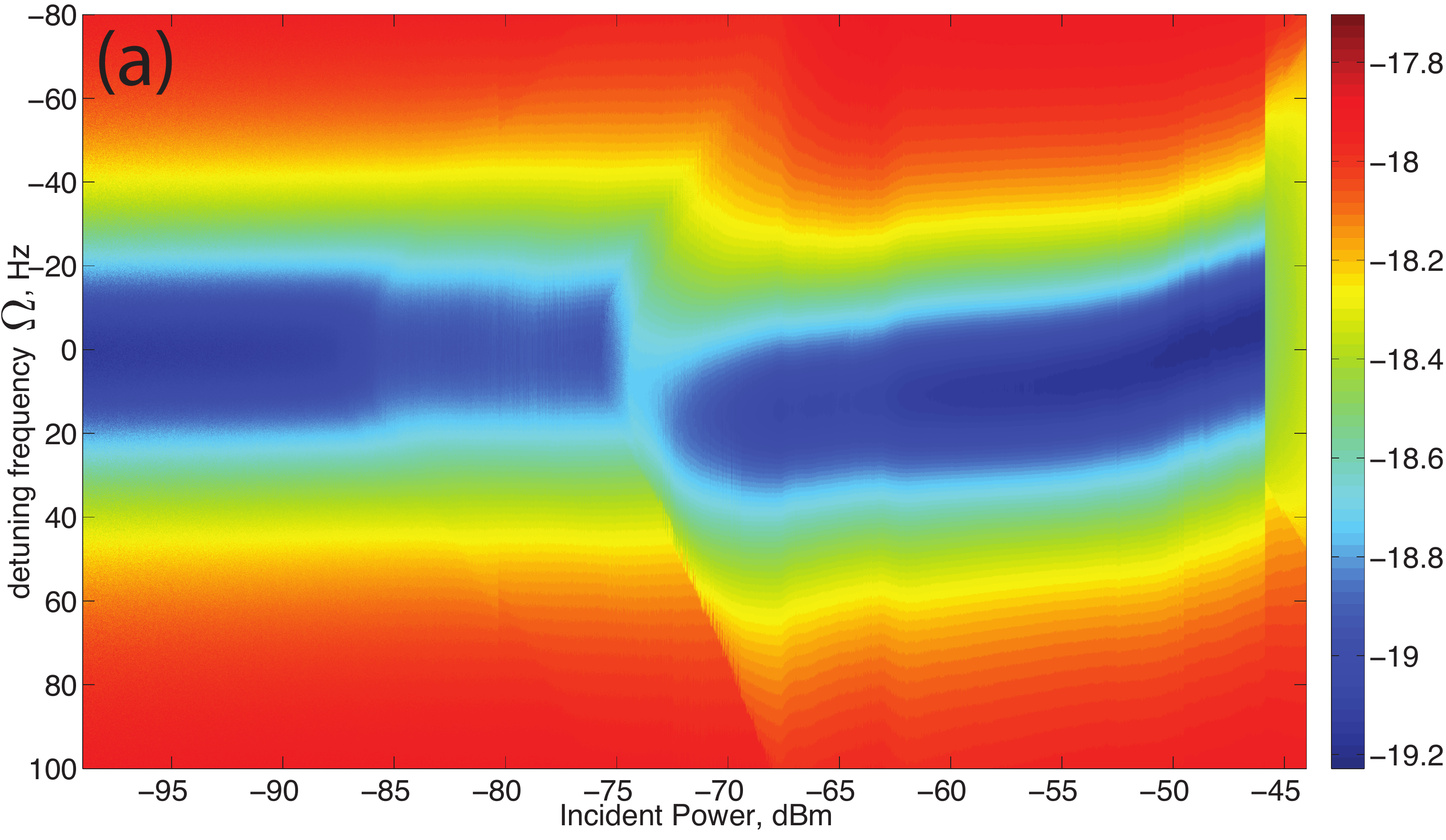}
                \label{magA5}
        \end{subfigure}%
        ~ 
        \begin{subfigure}[b]{0.5\textwidth}
                \centering
                \includegraphics[width=\textwidth]{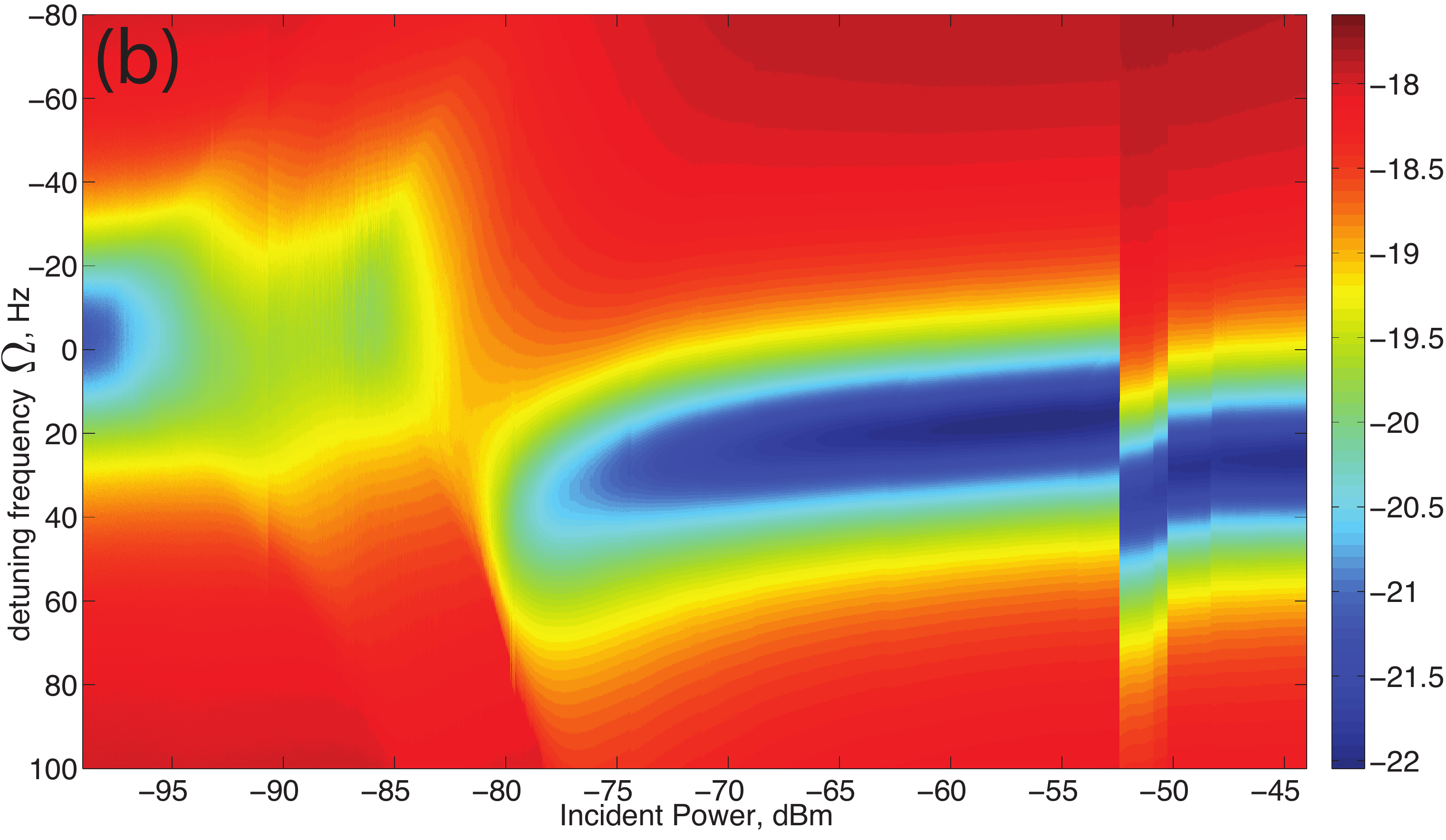}
                \label{mag5B}
        \end{subfigure}        
        ~
         \centering
        \begin{subfigure}[b]{0.5\textwidth}
                \centering
                \includegraphics[width=\textwidth]{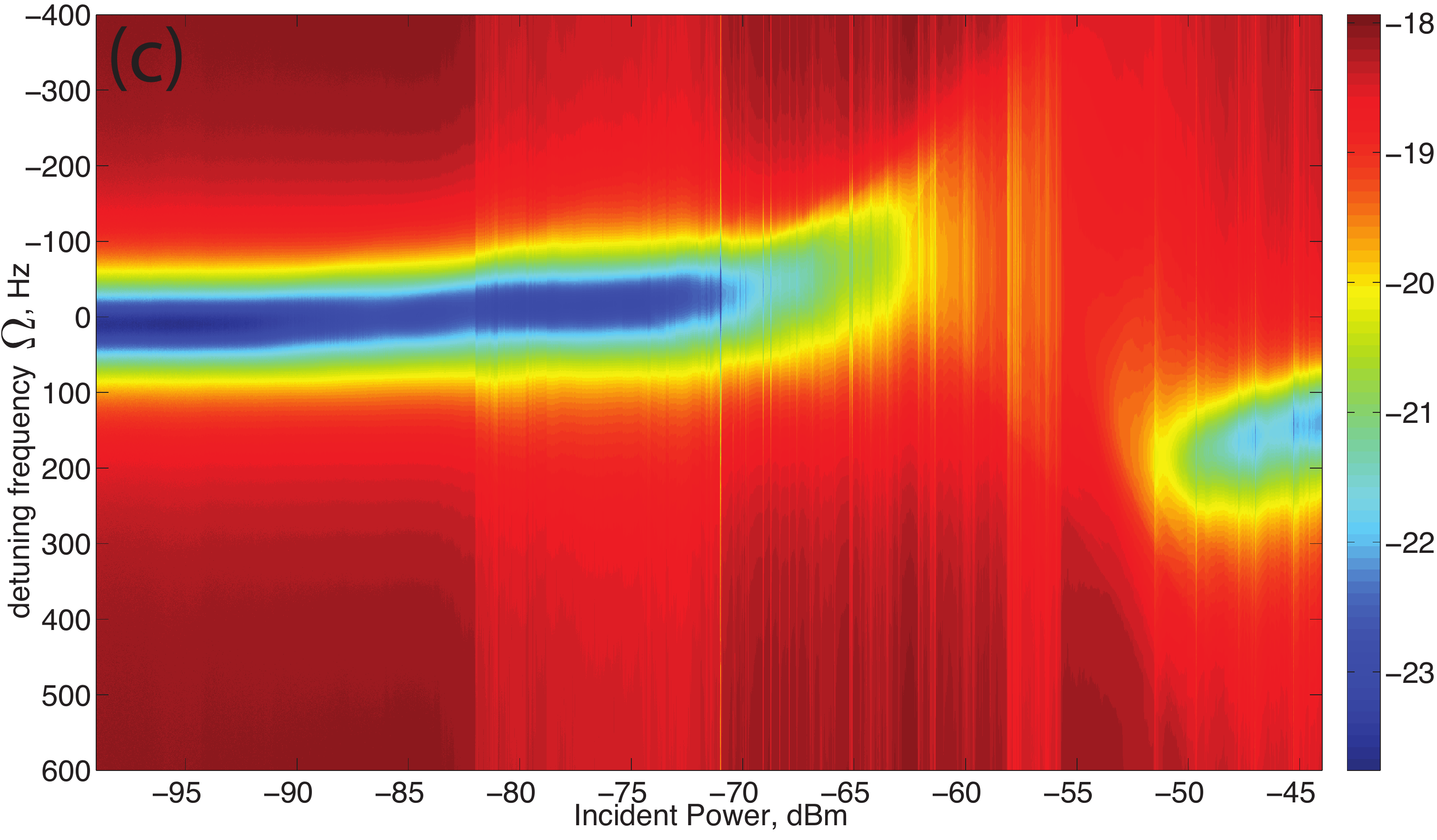}
                \label{mag5C}
        \end{subfigure}%
           ~ 
        \begin{subfigure}[b]{0.5\textwidth}
                \centering
                \includegraphics[width=\textwidth]{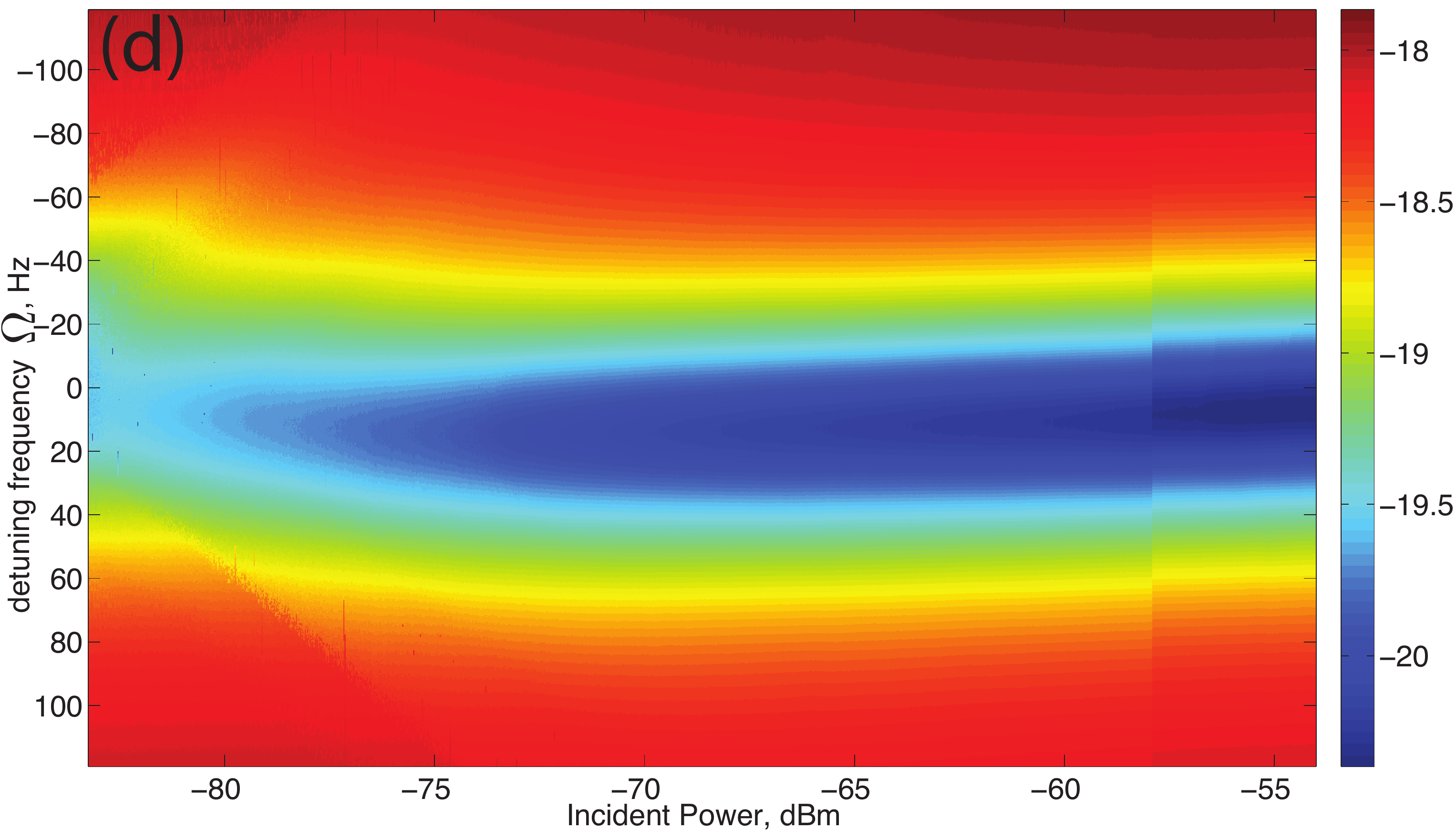}
               \label{mag5D}
        \end{subfigure}        
        \caption{Power Sensitivity of some anharmonics of the 5th OT (a: $99.91010$, b: $99.91792$, c: $99.97689$, d: $99.96408$) in terms of magnitude of the reflection coefficient. Here (a) and (d) show radially moving discontinuities, 'jumps', while (b) exhibits self-induced transparency and (b) and (c) demonstrate discontinuities in the power sweep.}\label{mags3}
\end{figure*}

\begin{figure}[t!]
\centering
\includegraphics[width=3.25in]{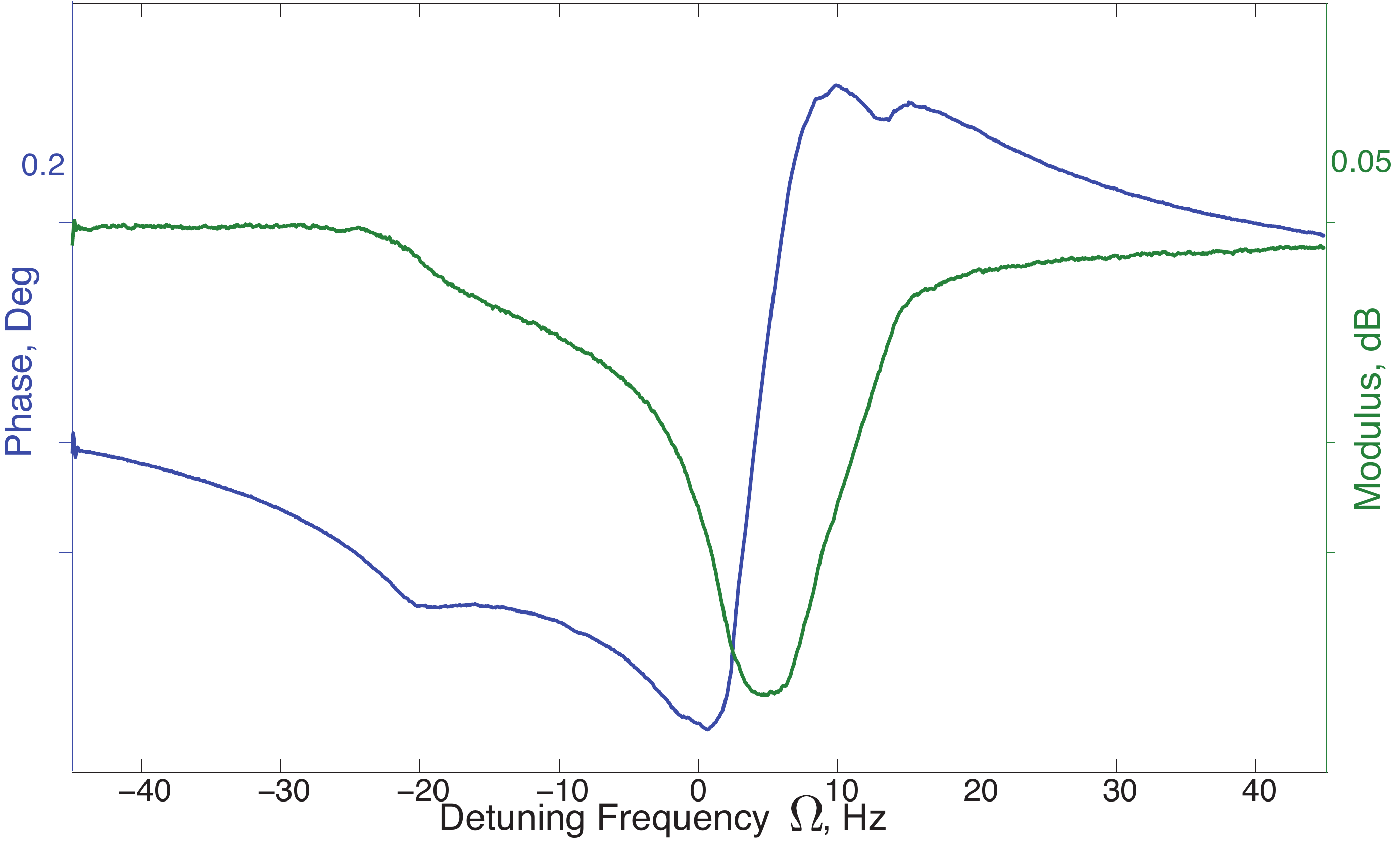}
  \caption{Phase and magnitude response of the device in the vicinity of an anharmonic of the 3rd OT ($59.94871$~MHz) at incident power of $-86.42$~dBm. Here we observed doubling of the frequency response extrema an effect of self-induced transparency.}\label{transparency-42p34}
\end{figure}

\begin{figure}[t!]
\centering
\includegraphics[width=3.25in]{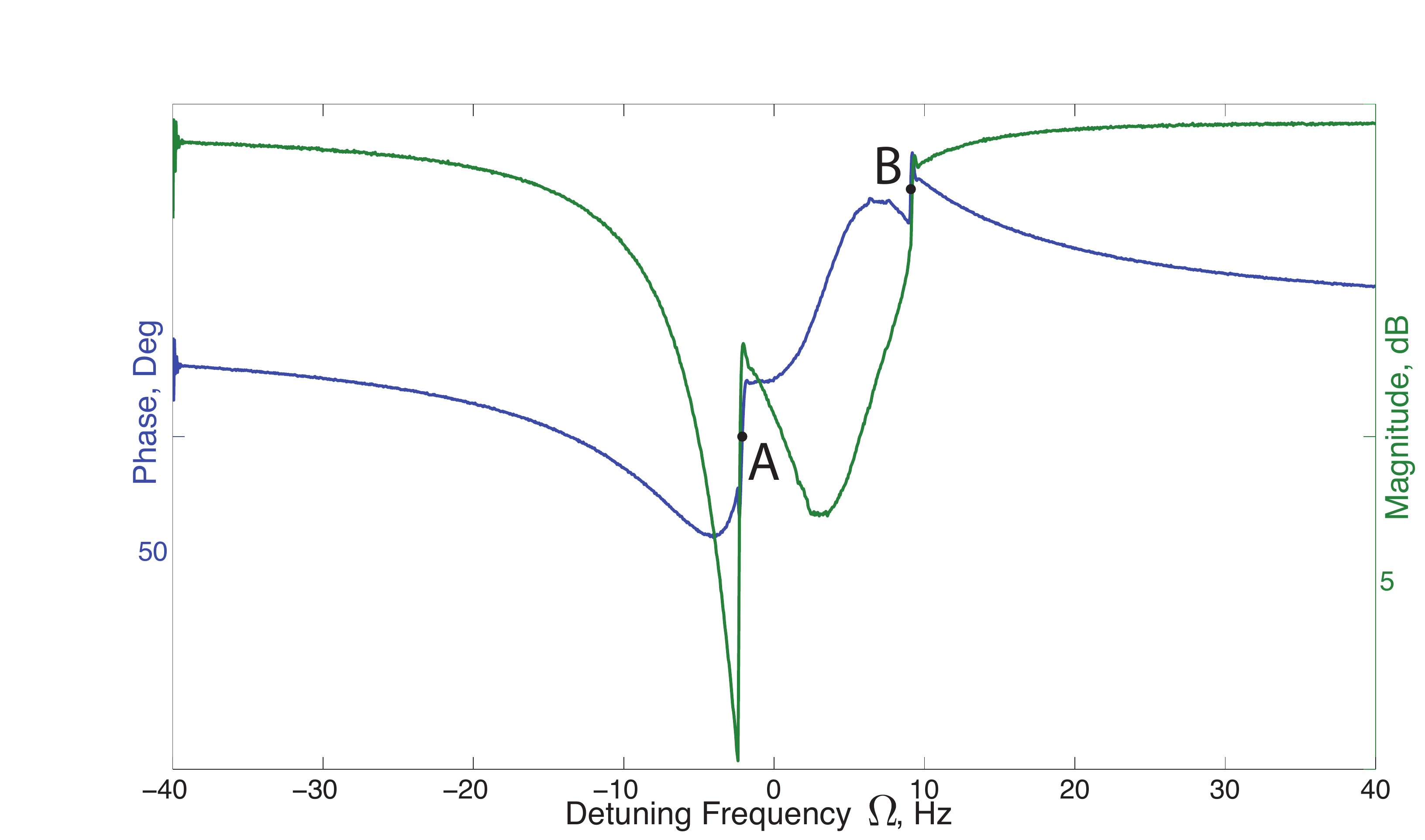}
 \caption{ Phase and magnitude response of the device in the vicinity of an anharmonic of the fundamental resonance ($19.98659$~MHz) at incident power of $-42.34$~dBm. Two discontinuity points A and B are at $-2.5$ and $9$~Hz.}\label{discont-86p42}
\end{figure}

\begin{figure}[t!]
\centering
\includegraphics[width=3.25in]{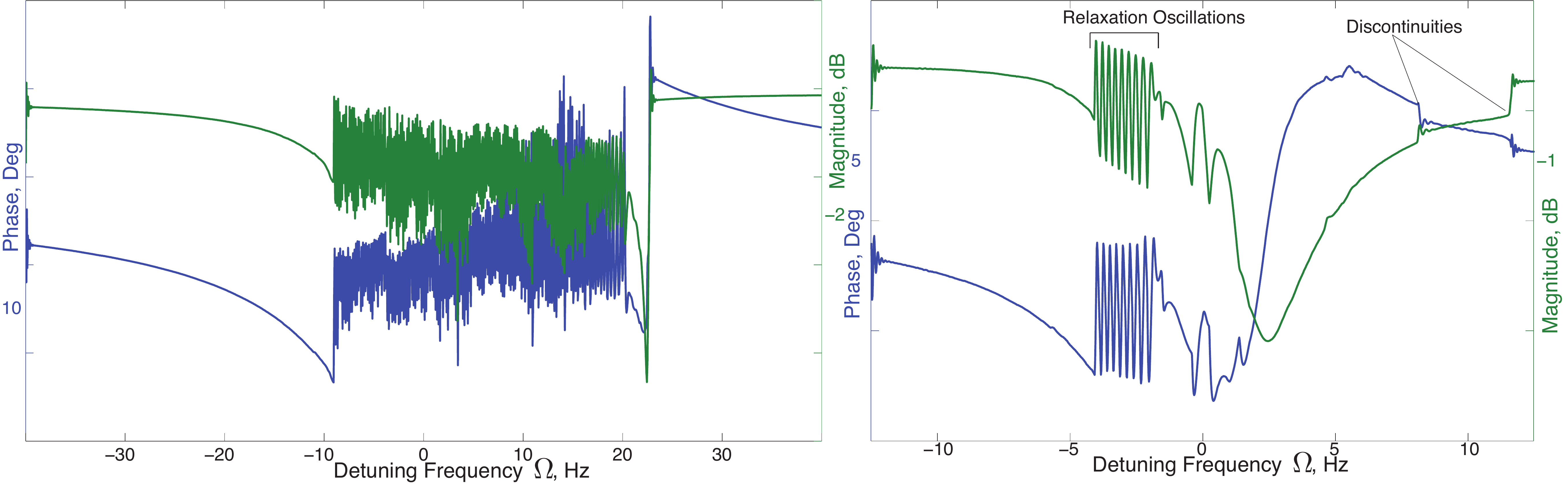}
 \caption{Phase and magnitude response of the device in the vicinity of anharmonics of the 3rd OT ($59.92972$~MHz) at incident power of $-51.3462$~dBm (left) and fundamental resonance ($20.04608$~MHz) at $-58.02$ dBm (right). The figure demonstrates instability region between $-8$ and $18$Hz represented by relaxation oscillations.}\label{oscillaitions-51p62}
\end{figure}

\begin{figure}[t!]
\centering
\includegraphics[width=3.25in]{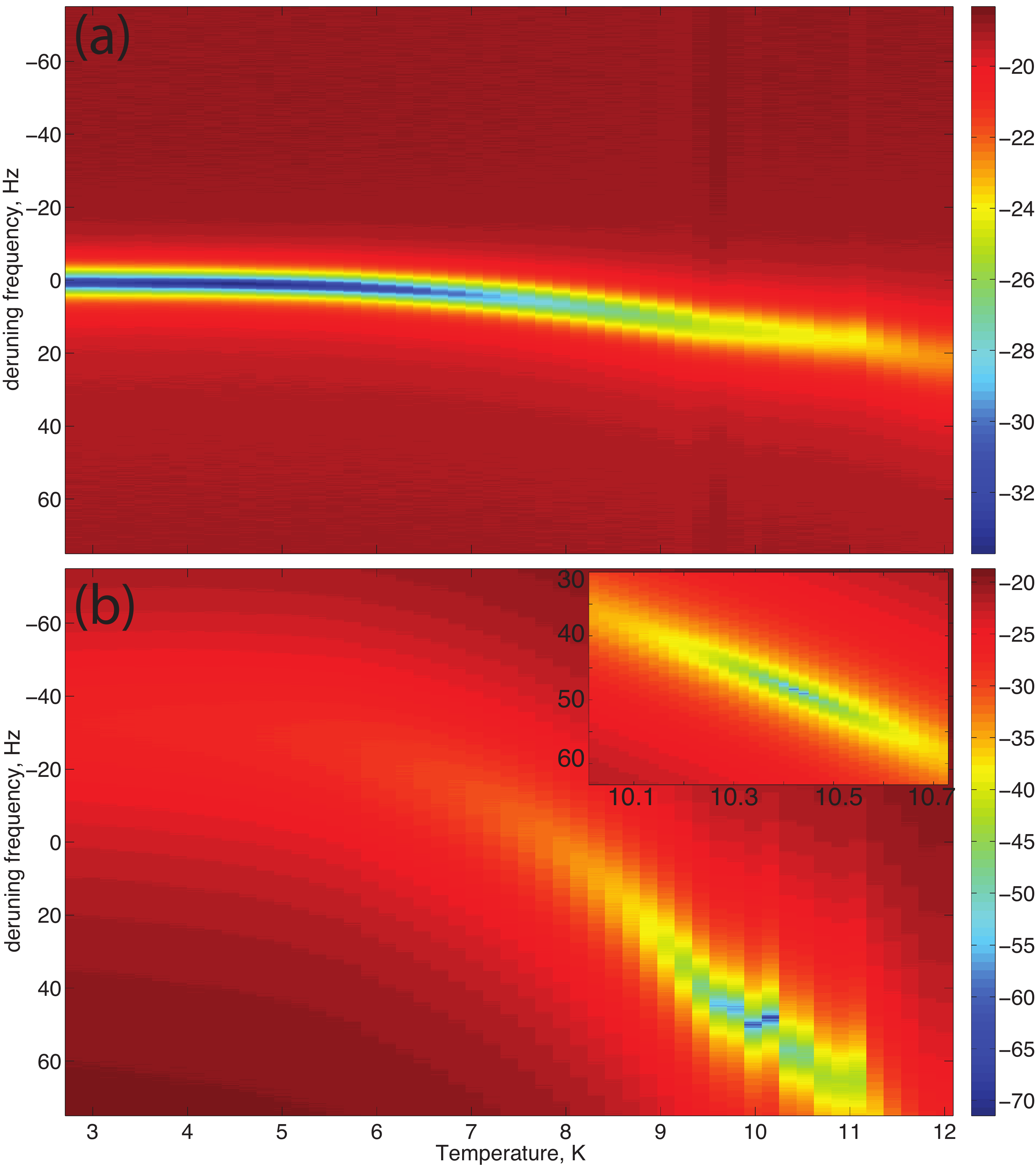}
  \caption{Temperature sensitivity of the two modes (a: $20.04608$~MHz, b: $99.9295$~MHz). The maximum mode coupling is achieved at $5$ and $10.45$K correspondingly. The inset shows zoomed in version of the plot near the reflection minimum. } \label{normalT}
\end{figure}

\begin{acknowledgments}
This work was supported by the Australian Research Council Grant No. CE110001013 and FL0992016. Authors want to thank Roman Boroditsky from NEL Frequency Controls for kindly supplying the samples. 
\end{acknowledgments}

\section*{References} 
%


\end{document}